\documentclass[letter,twocolumn]{jpsj3}
\usepackage{txfonts}
\usepackage{bm}
\usepackage{color}
\usepackage{amsmath}

\title{Topological Hall effect from strong to weak coupling}
\author{Kazuki Nakazawa$^1$\thanks{nakazawa@s.phys.nagoya-u.ac.jp}, 
Manuel Bibes$^2$ and 
Hiroshi Kohno$^1$}
\inst{$^1$Department of Physics, Nagoya University, Nagoya 464-8602, Japan \\ 
$^2$Unit\'e Mixte de Physique, CNRS, Thales, Universit\'e Paris-Saclay, 91767 Palaiseau, France} 
\abst{
Topological Hall effect (THE) of electrons coupled to a noncoplanar spin texture  
has been studied so far for the strong- and weak-coupling regimes separately; 
the former in terms of the Berry phase and the latter by perturbation theory. 
 In this letter,  we present a unified treatment  
in terms of spin gauge field  
by considering not only the adiabatic (Berry phase) component of the gauge field 
but also the nonadiabatic component.  
 While only the adiabatic contribution is important in the strong-coupling regime, 
it is completely canceled by a part of the nonadiabatic contribution in the weak-coupling regime, 
where the THE is governed by the rest of the nonadiabatic terms.
 We found a new weak-coupling region that cannot be accessed by a simple perturbation theory, 
where the Hall conductivity is proportional to $M$, with $2M$ being the exchange splitting 
of the electron spectrum. 
}
\begin{document}
\maketitle

 Berry phase\cite{Berry} is now recognized as an important viewpoint in condensed matter physics, 
connecting geometrical concepts and various physical phenomena.\cite{Niu} 
 Topological Hall effect (THE)\cite{Ye, Bruno} is  
one of the phenomena that are direct manifestations of the Berry phase. 
 Consider an electron moving in a smooth magnetization texture  
with an exchange coupling to it. 
 If the coupling is strong, the electron spin will adiabatically follow the texture and acquire a Berry phase. 
 Mathematically, this is described by a spin-dependent vector potential, $\sigma {\bm A}^z$, 
where $\sigma =\pm 1$ represents the spin direction, 
and induces a Hall effect. 
 This picture is valid if the electron stays in a given spin state $\sigma$ 
without experiencing spin-flip transitions.

 This adiabaticity condition fails in some cases. 
 When the exchange coupling is weak or the magnetic texture varies rapidly in space, 
electrons fail to adjust their spin to the local magnetization. 
 This is a nonadiabatic process that undermines the Berry phase picture. 
 However, it is known that THE exists even in the weak-coupling limit.\cite{Tatara, Nakazawa1}.   
 The condition when the adiabatic picture fails, especially in the diffusive regime 
(where the electrons feel the magnetic texture through their diffusive motion) 
is an important issue in quantum transport theory. 
 This question was first discussed in the context of conductance 
fluctuations,\cite{Stern, LKPB,LSG1, LSG2} and later for the THE\cite{Metalidis}. 
 In addition to the adiabaticity, the THE is characterized by the locality of the effective 
magnetic field; while the effective magnetic field is \lq\lq local'' in the Berry phase picture, 
it is \lq\lq nonlocal'' in the weak-coupling regime studied so far.\cite{Tatara, Nakazawa1} 
 This  \lq\lq locality condition'' is also important for THE.

 In this letter, we study the THE in both strong- and weak-coupling regimes in a single theoretical framework. 
 Focusing on the diffusive regime, we use the method of spin gauge field\cite{Korenman, Volovik, Bazaliy, Kohno} 
to calculate the adiabatic and nonadiabatic contributions to the topological Hall conductivity (THC). 
 A closely related analysis focussing on the weak-coupling regime will be presented in a longer 
paper,\cite{Nakazawa2} where the THE is analyzed by two other methods.

\begin{figure}[t]
 \begin{center}
  \includegraphics[width=80mm]{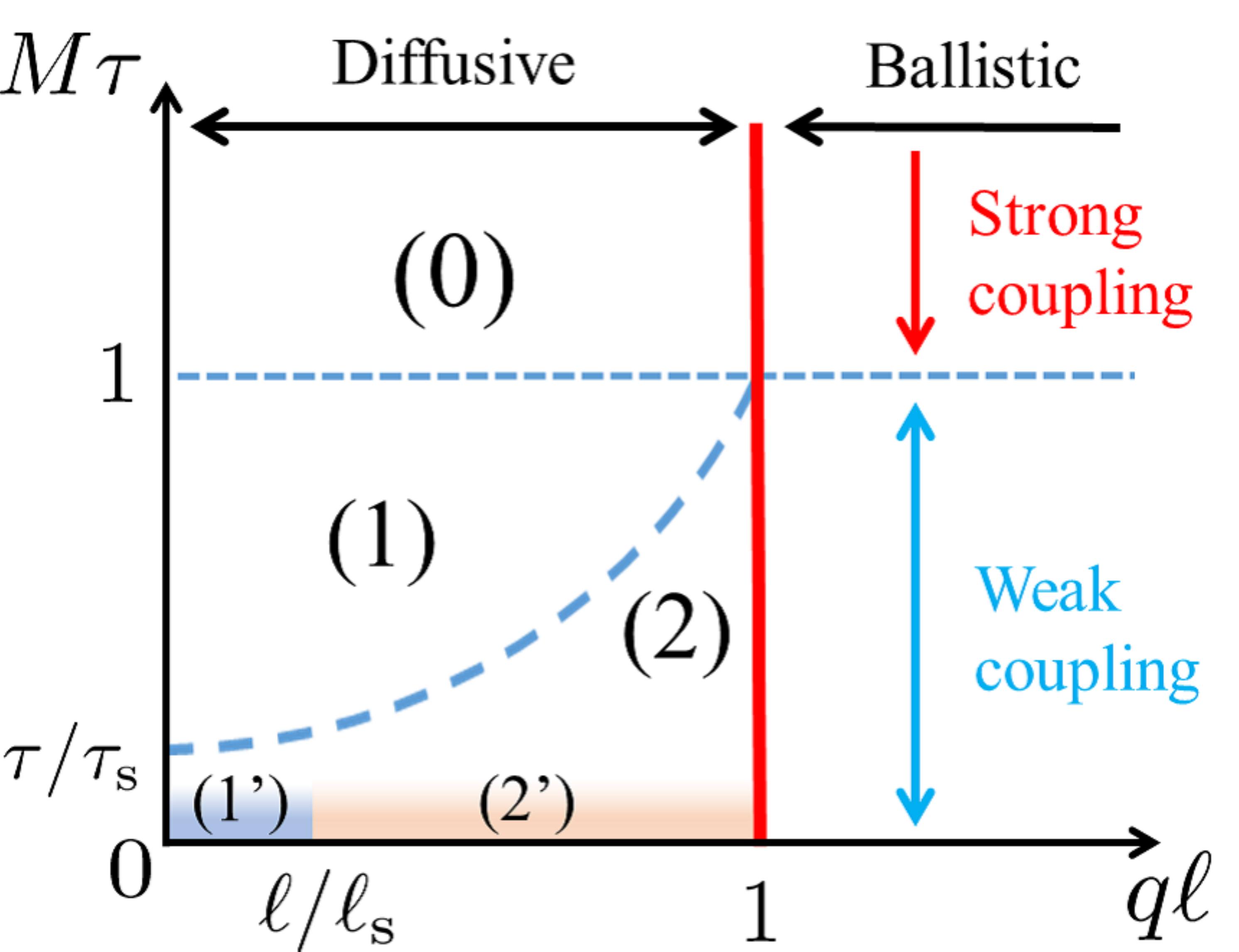}
 \end{center}
 \caption{(Color online) Characteristic regions in the diffusive regime $q\ell<1$ 
 in the plane of $M$ and $q$. 
 (0) Strong-coupling region. 
 (1) Weak-coupling and local-effective-field region. 
 (2) Weak-coupling and nonlocal-effective-field region.  
 Regions 0 and 1 are separated by the line $M \tau = 1$, 
and regions 1 and 2 by the parabola $M\tau = (q\ell)^2$. 
 In the presence of spin relaxation, the latter moves to $M\tau = (q\ell)^2 + 3 \tau/\tau_{\rm s}$, 
and there appear two more regions; ($1'$)  local, and ($2'$) nonlocal. 
 Note that $M\tau$ means $\tau/2\tau_{\rm ex}$ if one defines the \lq\lq exchange time'' 
$\tau_{\rm ex}=\hbar/2M$}. 
 \label{fig:34}
\end{figure}

 Several characteristic regions revealed in this study are summarized in Fig.~\ref{fig:34} 
in the plane of $M$ and $q$, where  $M$ is the exchange coupling constant and 
$q^{-1}$ is the characteristic length scale of the spin texture. 
 Both are made dimensionless with the electron scattering time $\tau$ 
and mean free path $\ell = v_{\rm F} \tau$ 
(by setting $\hbar = 1$). 
 In the absence of spin relaxation, the diffusive regime, i.e., $q\ell <1$, is classified into three regions, 
\begin{align}
 &{\rm Region \ 0} : \ (q\ell)^2 < 1  < M\tau ,  \nonumber \\    
 &{\rm Region \ 1} : \ (q\ell)^2 < M\tau < 1 ,  \nonumber \\ 
 &{\rm Region \ 2} : \ M\tau < (q\ell)^2  < 1 .  \nonumber 
\end{align}
 When the spin-relaxation time $\tau_{\rm s}$  
(or spin-diffusion length $\ell_{\rm s}$) is finite, 
there appear two more regions ($1'$ for $q \ell_{\rm s} < 1$, and $2'$ for $q \ell_{\rm s} > 1$) 
in the \lq\lq weakest-coupling'' region, $M \tau_{\rm s} <1$ (or $\tau_{\rm s} < \tau_{\rm ex}$). 
 We found that the adiabaticity holds for $M\tau > 1$ (region 0),\cite{Metalidis} 
and the locality of the effective magnetic field holds for $(q\ell)^2 < M\tau $ or $q\ell_{\rm s} < 1$ 
(regions 0, 1 and $1'$). 
 The expression of THC in each region, written as $\sigma_{xy}^{(n)}$ for region $n$, 
is given by eqs.~(\ref{eq:0}), (\ref{eq:1}), (\ref{eq:2}), (\ref{eq:1'}) and (\ref{eq:2'}) below.

 For an explicit analysis, we consider free electrons coupled to a continuous spin texture and subjected to impurity scattering. 
 The Hamiltonian is given by $H = H_0 + H_{sd}$,  
\begin{align}
 &H_0 
= \int d{\bm r} \, 
 c^{\dagger} ( {\bm r}) \left( - \frac{\nabla^2}{2m} - \epsilon_{\rm F} + V_{\rm imp} ({\bm r}) \right)  c ( {\bm r} )  , 
\label{eq:H0}
\\
 &H_{sd} 
= - M \int d^3r \ \bm{n}({\bm r}) \cdot ( c^{\dagger} ({\bm r}) \, \bm{\sigma} c ({\bm r}) )  , 
\label{eq:Hsd}
\end{align}
where $H_0$ describes the kinetic energy and random impurity potential, 
$V_{\rm imp} ({\bm r}) = u_{\rm i} \sum_j \delta ( {\bm r} - {\bm X}_j )$, 
and $H_{sd}$ is the exchange coupling to the spin texture ${\bm n}({\bm r})$. 
 Here, 
$c^{\dagger} = ( c^{\dagger}_{\uparrow}, c^{\dagger}_{\downarrow} )$ 
is an electron creation operator, 
${\bm \sigma}=( \sigma^{x}, \sigma^{y}, \sigma^{z} )$ are Pauli matrices, 
$u_{\rm i}$ and ${\bm X}_{i}$ are the potential strength and position, respectively, 
of normal impurities, 
and $M=J_{sd} S$ is the coupling constant of the exchange interaction $J_{sd}$ times 
the magnitude of the localized spin $S$. 
 $M$ is related to the \lq\lq exchange time'' $\tau_{\rm ex}$ by $2M=\hbar/\tau_{\rm ex}$. 
 In this letter, we study the Hall response of electrons under a given, static spin texture ${\bm n}({\bm r})$. 
%%%%%%%%%%%%%%%%%%%%%%%%%%%
%%%%%%%%%%%%%%%%%%%%%%%%%%%

 To treat the effects of texture, 
we move to the \lq\lq rotated frame" in which the spin quantization axis of the electrons 
is taken to be ${\bm n}({\bm r})$ at each point ${\bm r}$ of space. 
 Mathematically, this corresponds to diagonalizing $H_{sd}$ locally by transforming 
the electron spinor as $c({\bm r})= U({\bm r}) a({\bm r})$, where $U$ is an SU(2) matrix that satisfies 
$ U^{\dagger}({\bm r}) ( {\bm n}({\bm r}) \cdot {\bm \sigma} ) U({\bm r}) = \sigma^{z}$.
 As a price, there arises an SU(2) gauge potential (or gauge field),\cite{com_A} 
$ {\cal A}_{\mu} = -i U^{\dagger}({\bm r}) \partial_{\mu} U({\bm r}) $. 
 This is a $2 \times 2$ matrix in spin space, 
\begin{align}
  {\cal A}_{\mu} 
 =  A_{\mu}^{\alpha} \frac{\sigma^{\alpha} }{2}
 = A_{\mu}^{z} \frac{\sigma^{z}}{2} + {\bm A}_{\mu}^{\perp} \cdot \frac{{\bm \sigma}^{\perp}}{2}  , 
\label{eq:A1}
\end{align}
where the upper and lower indices of $A_{\mu}^{\alpha}$ represent spin and real-space 
components, respectively, and ${\bm A}_{\mu}^{\perp} = (A_\mu^x , A_\mu^y, 0)$, 
${\bm \sigma}^{\perp} = (\sigma^x , \sigma^y, 0)$.  
 The diagonal component $A_{\mu}^{z}$ preserves the spin state and describes adiabatic processes, 
whereas the off-diagonal component ${\bm A}_{\mu}^{\perp}$ induces a spin-flip transition and 
describes nonadiabatic processes. 
 The former acts as a spin-dependent vector potential  
and gives a mathematical expression of the Berry phase.

 In the rotated frame, the Hamiltonian is given by 
\begin{align}
 {\cal H}  
= \int d{\bm r} \, 
 a^{\dagger} ( {\bm r}) \left( - \frac{(\nabla + i {\cal A})^2}{2m} - \epsilon_{\rm F} 
    - M \sigma^z + V_{\rm imp} \right)  a ( {\bm r} ) . 
\end{align}
 We treat the gauge field perturbatively. 
 This is justified when the spatial variation of magnetic texture 
is slow ($q\ell \ll 1$) since the gauge field involves a spatial derivative of the spin texture, 
and is small when the texture varies slowly. 
 Note that there is no restriction on the magnitude of $M$, allowing us to study the strong- and 
weak-coupling regimes in this single framework.

 In the Berry phase picture of THE,\cite{Bruno} 
the gauge field acts on the electrons via an effective (spin-dependent) magnetic field, 
\begin{align}
 B_{{\rm s}, z} 
&= (\nabla \times {\bm A}^{z} )_z 
 =  {\bm n} \cdot ( \partial_{x} {\bm n} \times \partial_{y} {\bm n} )  . 
\label{eq:B_Az}
\end{align}
 Interestingly, the nonadiabatic component ${\bm A}_i^{\perp} $ produces the same field 
(hence using the same notation $B_{{\rm s}, z}$),  
\begin{align}
 B_{{\rm s}, z} 
&=  ( {\bm A}_{x}^{\perp} \times {\bm A}_{y}^{\perp} )^z 
 = {\bm n} \cdot ( \partial_{x} {\bm n} \times \partial_{y} {\bm n} )  . 
\label{eq:B_Aperp}
\end{align}
 The equality of the two 
comes from the fact that the SU(2) gauge field (\ref{eq:A1}) 
arises from a pure gauge transformation, and the SU(2) field strength vanishes, 
${\cal F}_{ij} = -i \, [ \partial_i + i{\cal A}_i  , \partial_j + i{\cal A}_j ] = 0$. 
 In terms of the components, 
${\cal F}_{ij} = F_{ij}^\alpha \, \sigma^\alpha/2$, this is expressed as\cite{com_A,Shibata} 
\begin{align}
 F_{ij}^\alpha  &=  \partial_i  A_j^\alpha - \partial_j A_i^\alpha - ({\bm A}_i \times {\bm A}_j)^\alpha  
 = 0 . 
\label{eq:F_ij^a}
\end{align}
 The presence of the magnetization breaks the SU(2) gauge symmetry 
down to U(1), and the \lq\lq projected'' U(1) gauge field ${\bm A}^z$ can have finite field strength, 
which is eq.~(\ref{eq:B_Az}). 
 In the weak-coupling regime, it is known that the effective field can be \lq\lq nonlocal'', 
i.e., it is related to the spin texture in a nonlocal way.\cite{Tatara, Nakazawa1}. 
 This is also the case in the present case of continuous texture, as we will see 
[eqs.~(\ref{eq:2})-(\ref{eq:2_d}) and (\ref{eq:2'})-(\ref{eq:2'_d})]. 
%%%%%%%%%%%%%%%%%%%%%%%%%%%%%%
%%%%%%%%%%%%%%%%%%%%%%%%%%%%%%

 To calculate THC, we use Kubo formula for electrical conductivity, 
\begin{align}
\sigma_{ij} ({\bm Q},\omega)
&= \frac{K_{ij}^{\rm R}({\bm Q}, \omega) - K_{ij}^{\rm R}({\bm Q}, 0)}{i\omega}
, \\
K_{ij}^{\rm R}({\bm Q}, \omega) 
&= i \int_{0}^{\infty} dt \ {\rm e}^{i(\omega+i0)t} 
     \left< \left[ J_{i}({\bm Q},t), J_{j}({\bm 0},0) \right] \right> , 
\end{align}
and extract the antisymmetric part, $ \frac{1}{2} ( \sigma_{xy} -\sigma_{yx} ) \equiv \sigma_{xy}^{\rm H}$. 
(We will suppress the superscript H, however.)
 Here we are looking at the Fourier component of the electric current, 
\begin{align}
 J_{i} ({\bm Q}) = - e \sum_{\bm k}
  v_i \, a_{{\bm k}-{\bm Q}/2}^{\dagger} a_{{\bm  k} + {\bm Q}/2} 
  - \frac{e}{2m} \sum_{{\bm k}, {\bm q}}  A_{i}^{\alpha} ({\bm q}) a_{{\bm k} + {\bm q}}^{\dagger} 
   \sigma^{\alpha} a_{{\bm k} + {\bm Q}}  , 
\end{align}
in response to a spatially uniform electric field. 
 Here ${\bm k}$ is a wave vector, $v_i = \hbar k_i/m$, and $A_{i}^{\alpha}({\bm q})$ is the 
Fourier component of the gauge field. 
 The wave-vector ${\bm Q}$ of the current density comes from the spin texture (gauge field), 
and  $\omega$ is the frequency of the applied electric field. 
 The d.c.~THC of a macroscopic system is obtained 
by taking the limit ${\bm Q} \to {\bm 0}$ first, and then setting $\omega \to 0$.

\begin{figure}[t]
 \begin{center}
  \includegraphics[width=80mm]{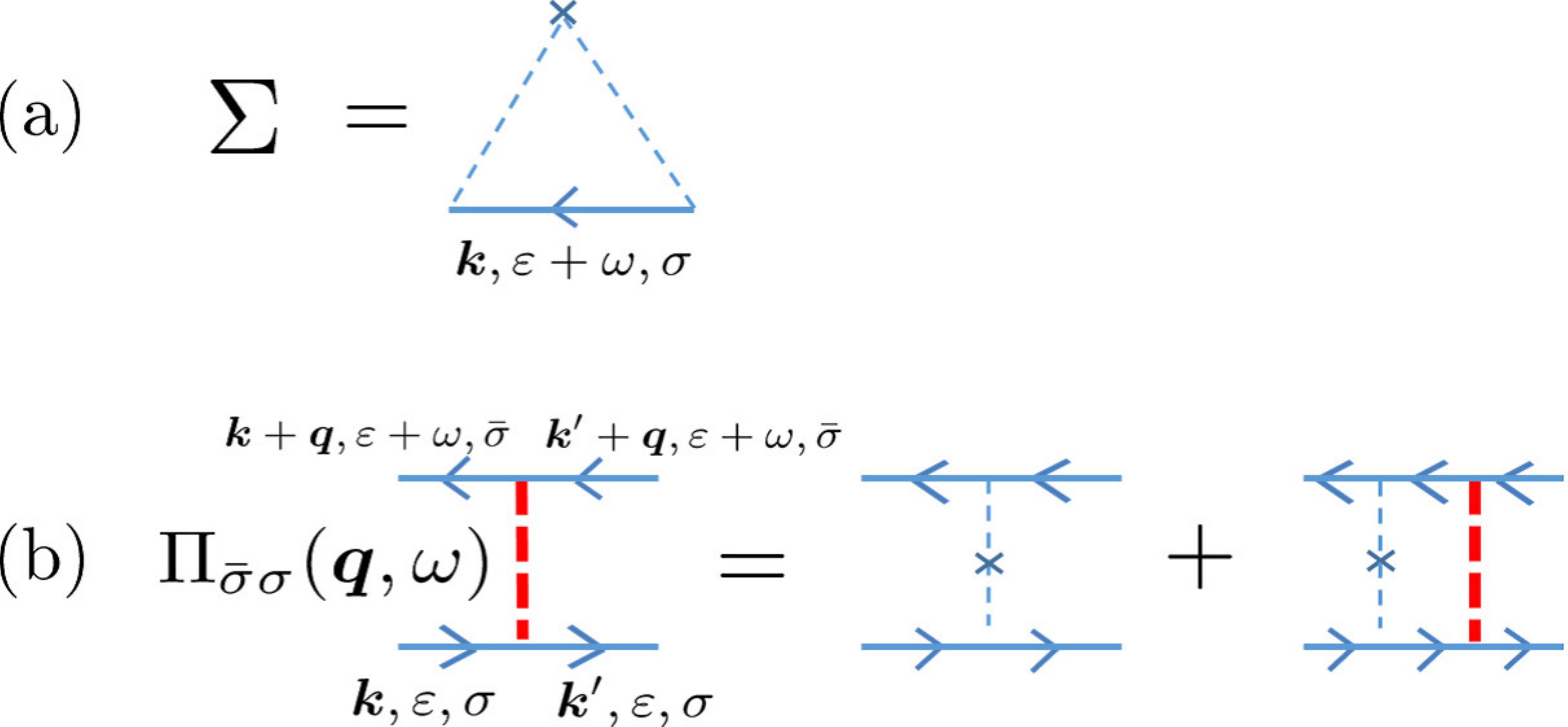}
\end{center}
 \caption{(Color online) 
 Diagrammatic expression of self-energy and vertex corrections due to random impurities. 
 The solid lines are Green functions of electrons, and 
 the thin broken line with a cross represents impurity scattering. 
 (a) Self-energy in the Born approximation. 
 (b) Diffusion propagator, $\Pi_{\bar\sigma \sigma} ({\bm q},\omega)$, in the ladder approximation. 
 The upper (lower) line represents retarded (advanced) Green function [Eq.~(\ref{eq:green})]. 
}
 \label{fig:diffself}
\end{figure}

 With the self-energy evaluated in the Born approximation (Fig.~\ref{fig:diffself} (a)), 
the Green function in the rotated frame is given by
$G_{\bm k \sigma}^{\rm R(A)} (\varepsilon) 
= (\varepsilon - \varepsilon_{\bm k} + \sigma M \pm i/2\tau_{\sigma})^{-1}$, 
where 
$\tau_{\sigma} = (2\pi n_{\rm i} u_{\rm i}^2 \nu_{\sigma})^{-1}$ is the scattering time, 
${\nu}_{\sigma}$ is the density of states at the Fermi energy of electrons with 
spin $\sigma$, and $n_{\rm i}$ is the impurity concentration. 
 In response functions, we consider ladder-type vertex corrections, 
whose essential ingredient is the spin diffusion propagator (Fig.~\ref{fig:diffself} (b)), 
\begin{align}
  \Pi_{\bar\sigma \sigma} ({\bm q},\omega) 
&= \frac{1}{2 \pi\nu \tau^2} \,  
    \frac{1 + 2i\sigma M \tau }{Dq^2 + 2i\sigma M - i\omega + \tau_{\rm s}^{-1} } . 
\label{eq:mVC}
\end{align}
 Here we assumed $q\ell < 1$ and $M\tau < 1$, 
with $q$ being the wave-vector of the texture, 
$\ell = v_{\rm F} \tau$ the electron mean free path, and 
$D = \frac{1}{3}v_{{\rm F} }^{2} \tau$ the diffusion constant. 
 This vertex correction is relevant only in the weak-coupling regime,\cite{SM} 
hence we dropped unimportant spin dependence in $D_\sigma$ and $\tau_\sigma$.   
 Physically, Eq.~(\ref{eq:mVC}) describes diffusion ($\sim Dq^2$), precession ($\sim 2i\sigma M$) 
 and relaxation ($\sim \tau_{\rm s}^{-1}$) 
of transverse spin density of electrons. 
 We introduced the spin-relaxation time $\tau_{\rm s}$ by hand.

\begin{figure}[t]
 \begin{center}
  \includegraphics[width=85mm]{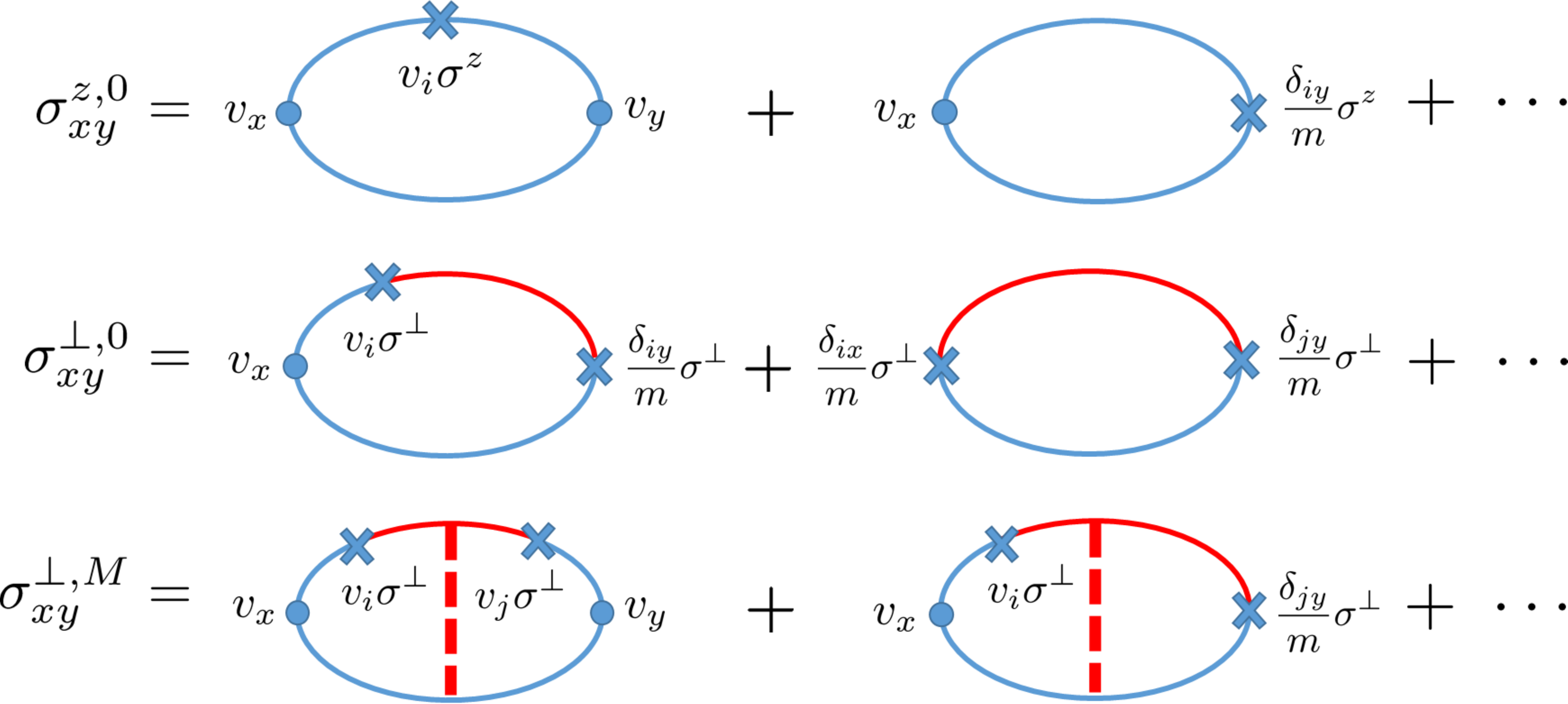}
 \end{center}
 \caption{(Color online) Feynman diagrams for the THC in the gauge-field method. 
  The first line ($\sigma_{xy}^{z,0}$) is the adiabatic contribution. 
  The second ($\sigma_{xy}^{\perp,0}$) and third ($\sigma_{xy}^{\perp, M}$) lines represent 
  nonadiabatic contributions without and with vertex corrections, respectively.  
 The blue thick cross with $v_i \sigma^{\alpha}$ represents the coupling to the gauge field $ A_i^{\alpha} $. 
 The blue (red) solid line is the Green function of electrons with spin $\sigma$ ($\bar{\sigma}$). 
}
 \label{fig:THCA}
\end{figure}

 We now demonstrate that, in the present formulation based on the spin gauge field, 
the Feynman diagrams shown in Fig.~\ref{fig:THCA} describe THC in all regions. 
 They are classified into the adiabatic ($\sigma_{xy}^{z}$) and the nonadiabatic ($\sigma_{xy}^{\perp}$) terms. 
 Details of the calculation are described in the Supplemental material.\cite{SM}

 In the strong-coupling regime ($M\tau > 1$, region 0), 
the adiabatic and nonadiabatic contributions are calculated as\cite{Fukuyama,SM}  
\begin{align}
  \sigma_{xy}^{z, 0} ({\bm q}, \omega) 
&= -\frac{e^2}{2m} \left( i{\bm q} \times {\bm A}^{z} ({\bm q}) \right)_{z} \sum_{\sigma} \sigma D_{\sigma} \nu_{\sigma} \tau_{\sigma} , 
\label{eq:s_xy^z0_R0}
\\ 
  \sigma_{xy}^{\perp} ({\bm Q}, \omega) 
&= \sigma_{xy}^{\perp, 0} + \sigma_{xy}^{\perp, M} 
= -\frac{1}{72} \left(  \frac{e}{m} \right)^2 \frac{\nu}{M} 
            \bigl[ {\bm A}_{x}^{\perp} ({\bm q}) \times {\bm A}_{y}^{\perp} ({\bm q}') \bigr]^{z}  , 
\label{eq:s_xy^p_R0}
\end{align}
where ${\bm Q} = {\bm q} + {\bm q}'$. 
 The latter is independent of the scattering time $\tau$ and smaller than the former  
by a factor of $(M \epsilon_{\rm F} \tau^2)^{-1}$. 
 Therefore, the Berry phase picture holds in region 0, and the total THC is given by 
($\langle \cdots \rangle$ means spatial average) 
\begin{align}
  \sigma_{xy}^{(0)} \simeq \sigma_{xy}^{z} 
&= -\frac{e^2}{2m} \langle B_{{\rm s}, z} \rangle \sum_{\sigma} \sigma D_{\sigma} \nu_{\sigma} \tau_{\sigma} 
, 
\label{eq:0}
\end{align}
in agreement with Bruno {\it et al.}\cite{Bruno}.

 In the weak-coupling regime ($M\tau < 1$), each contribution in Fig.~\ref{fig:THCA} 
is calculated as\cite{Fukuyama} 
\begin{align}
 \sigma_{xy}^{z,0} ({\bm q}, \omega) 
&= -\frac{1}{3} \left( \frac{e}{m} \right)^2 \nu M \tau^2 
    \left( i{\bm q} \times {\bm A}^{z} ({\bm q}) \right)_{z} , 
\label{eq:s_xy^z0}
\\
 \sigma_{xy}^{\perp,0} ({\bm Q}, \omega) 
&= +\frac{1}{3} \left( \frac{e}{m} \right)^2 \nu M \tau^2  \, 
 \bigl[ {\bm A}_{x}^{\perp} ({\bm q}) \times {\bm A}_{y}^{\perp} ({\bm q}') \bigr]^z  , 
\label{eq:s_xy^p0}
\\
 \sigma_{xy}^{\perp, M} ({\bm Q}, \omega) 
&= -\frac{4}{9} \left( \frac{e}{m} \right)^2 \nu M^3 \tau^4 \, 
     \left[  |\Gamma (q)|^2 + |\Gamma (q')|^2 \right]  
\nonumber \\
& \times 
  \bigl[ {\bm A}_{x}^{\perp} ({\bm q}) \times {\bm A}_{y}^{\perp} ({\bm q}') 
                + {\bm A}_{x}^{\perp} ({\bm q}') \times {\bm A}_{y}^{\perp} ({\bm q}) \bigr]^z , 
\label{eq:s_xy^pM}
\end{align}
where $|\Gamma (q)|^2 + |\Gamma (q')|^2$, with 
\begin{align}
 \Gamma (q) &=  \frac{1}{ ( Dq^2  + \tau_{\rm s}^{-1}  + 2i |M| ) \tau} , 
\end{align}
comes from the vertex correction. 
 Because of the weak-coupling condition $M\tau < 1$, 
we retained only the low-order terms with respect to $M$ 
(dropping the spin dependence in $\nu_\sigma$, etc.), 
except for the diffusion propagator in eq.~(\ref{eq:s_xy^pM}), 
which is \lq\lq singular'' having $M$ in the denominator.

 The adiabatic contribution $\sigma_{xy}^{z,0}$ [eq.~(\ref{eq:s_xy^z0})] can be obtained 
from eq.~(\ref{eq:s_xy^z0_R0}) by retaining the lowest order terms in $M$.  
 Surprisingly, it is completely canceled by the nonadiabatic contribution, 
$\sigma_{xy}^{\perp ,0}$ [eq.~(\ref{eq:s_xy^p0})], because of the relation (\ref{eq:F_ij^a}). 
 This cancellation occurs among the diagrams without vertex corrections, and seems robust 
as we will discuss below. 
 Therefore, the only relevant contribution is $\sigma_{xy}^{\perp, M}$ [eq.~(\ref{eq:s_xy^pM})], 
the nonadiabatic contribution with vertex corrections. 
 This means that  
 the adiabaticity condition is given by $M\tau > 1$, 
which agrees with the conclusion of Ref.~\cite{Metalidis} based on numerical methods.

 The behaviour of THC in the weak-coupling ($M\tau <1$) diffusive ($q\ell <1$) regime 
is thus determined by the spin diffusion propagator $\Gamma (q)$. 
 The classification 
shown in Fig.~1  
  is based on this observation.
 For a closer study, it is convenient to look at the real-space form,\cite{SM}  
\begin{align}
 \sigma_{xy}^{\perp, M} 
&= - \frac{8}{9} \left( \frac{e}{m} \right)^2 \nu 
    M^3 \tau^4 \, 
   {\rm Re} \left[ \langle {\bm n} \cdot ( \tilde {\bm d}_{x} \times \tilde {\bm d}_{y}^* ) \rangle \right] ,
\label{eq:12_r}
\\ 
 \tilde {\bm d}_{i} ({\bm r})  
&=   \frac{1}{4\pi D \tau } 
     \int d{\bm r}' \ 
     \frac{e^{-(a+ib) |{\bm r} - {\bm r}'|} }{ |{\bm r} - {\bm r}'| } \, 
    {\cal R}({\bm r}) \, {\cal R}^{-1} ({\bm r}') \, \partial_{i} {\bm n} ({\bm r}') , 
\label{eq:d}
\end{align}
where $a= \bigl[ \bigl( \sqrt{ \ell_{\rm s}^{-4} + \lambda^{-4} } + \ell_{\rm s}^{-2} \bigr)/2\bigr]^{1/2}$, 
$b = (2a\lambda^2)^{-1}$, 
$\ell_{\rm s} = \sqrt{D\tau_{\rm s}}$ is the spin-diffusion length as before, and 
$\lambda = \sqrt{\hbar D/2|M|} = \sqrt{D \tau_{\rm ex}} $ 
is the \lq\lq spin-precession length''. 
 $\cal R$ is an SO(3) matrix that relates the rotated to the original frame, ${\cal R} \hat{z} = {\bm n}$, 
 and satisfies\cite{Kohno} 
${\cal R} {\bm A}^\perp_i = - {\bm n} \times \partial_i {\bm n}$. 
 Because of the $q$-dependence of $\Gamma (q)$, $\tilde {\bm d}_{i} ({\bm r})$ is related to 
$\partial_{i} {\bm n} ({\bm r}') $ in a nonlocal way. 
 However, if the spin texture varies slowly compared to $a^{-1}$, this relation becomes a local one.  
 Each case is analyzed as follows. 

\noindent
$\bullet$ Weak spin relaxation, $\ell_{\rm s} \gg \lambda$ ($M \tau_{\rm s} \gg 1$, i.e., $\tau_{\rm s}/\tau_{\rm ex} \gg 1$)

\noindent
\  - Region 1 ($q \lambda < 1$, local) : $\tilde {\bm d}_i = (2iM\tau )^{-1} \partial_i {\bm n}$,   
\begin{align}
 \sigma_{xy}^{(1)} 
&= -\frac{4}{9} \left( \frac{e}{m} \right)^2 \langle B_{{\rm s}, z} \rangle \, \nu M \tau^2 .
\label{eq:1}
\end{align}

\noindent
\  - Region 2 ($q \lambda > 1$, nonlocal) :  
\begin{align}
 \sigma_{xy}^{(2)} 
&= - \frac{4}{9} \left( \frac{e}{m} \right)^2 \nu M \tau^2  \, 
   {\rm Re} \left[ \langle {\bm n} \cdot ( {\bm d}_{x}^{(2)} \times {\bm d}_{y}^{(2) \, *} ) \rangle \right] ,
\label{eq:2}
\\ 
 {\bm d}_{i}^{(2)} ({\bm r})  
&= \frac{1}{4\pi \lambda^2}  \int d{\bm r}' \ 
     \frac{e^{- (1+i)|{\bm r} - {\bm r}'|/\sqrt{2}\lambda}}{ |{\bm r} - {\bm r}'|} \, 
    {\cal R}({\bm r}) \, {\cal R}^{-1} ({\bm r}') \, \partial_{i} {\bm n} ({\bm r}')  . 
\label{eq:2_d}
\end{align}

\noindent
$\bullet$ Strong spin relaxation, $\ell_{\rm s} < \lambda$ ($M \tau_{\rm s} < 1 $, i.e., $\tau_{\rm s}/\tau_{\rm ex} < 1$)

\noindent
\  - Region $1'$ ($q \ell_{\rm s} < 1$, local) : $\tilde {\bm d}_i = (\tau_{\rm s}/\tau) \, \partial_i {\bm n}$,  
\begin{align}
 \sigma_{xy}^{(1')} 
&= - \frac{16}{9} \left( \frac{e}{m} \right)^2 \langle B_{{\rm s}, z} \rangle \, \nu M^3 \tau^2 \tau_{\rm s}^2 . 
\label{eq:1'}
\end{align}

\noindent
\  - Region $2'$ ($q \ell_{\rm s} > 1$, nonlocal) : 
\begin{align}
 \sigma_{xy}^{(2')}
&= - \frac{16}{9} \left( \frac{e}{m} \right)^2 \nu M^3 \tau^2  \tau_{\rm s}^2 \, 
       \langle {\bm n} \cdot ( {\bm d}_{x}^{(2')} \times {\bm d}_{y}^{(2')} ) \rangle ,
\label{eq:2'}
\\ 
 {\bm d}_{i}^{(2')} ({\bm r})  
&= \frac{1}{4\pi \ell_{\rm s}^2}  \int d{\bm r}' \ 
     \frac{e^{- |{\bm r} - {\bm r}'|/\ell_{\rm s}}}{|{\bm r} - {\bm r}'|} \, 
    {\cal R}({\bm r}) \, {\cal R}^{-1} ({\bm r}') \, \partial_{i} {\bm n} ({\bm r}')  . 
\label{eq:2'_d}
\end{align}

 The $M^3$-behaviour of THC in regions $1'$ and $2'$ is common in the perturbative regime,\cite{Tatara,Nakazawa1}  
reflecting the fact that a noncoplanar spin structure requires at least three spins. 
 The behaviour in region 2 is complex because of the nonlocal nature.

 The most striking result in this report is the $M$-linear behaviour of THC in region 1. 
 This is not obtained perturbatively. 
 In fact, the perturbative $M$-linear terms canceled out, $\sigma_{xy}^{z,0} + \sigma_{xy}^{\perp,0} =0$ 
[eqs.~(\ref{eq:s_xy^z0}) and (\ref{eq:s_xy^p0})], as we have seen. 
 This cancellation seems robust, ensured by the underlying SU(2) symmetry. 
 To see this, let us consider a perturbative expansion of THC in terms of the SU(2) gauge field, $A_i^\alpha$, 
and the magnetization ${\bm M} = M  \hat z$ (in the rotated frame).   
 Because of gauge invariance, it will start as\cite{com2} 
\begin{align}
 \sigma_{ij}^{\rm H} = c F_{ij}^\alpha M^\alpha + \cdots , 
\label{eq:s_ij=F_ij}
\end{align}
where $F_{ij}^\alpha$ is the SU(2) field strength [see the first equality in eq.~(\ref{eq:F_ij^a})], 
and $c$ is a coefficient. 
 This is similar to the Zeeman term, with ${\bm M}$ being a symmetry-breaking field. 
 Note that the \lq\lq coefficient'' $cF_{ij}^\alpha$ of the $M$-linear term is determined in the absence of $M$, 
hence it should reflect the full SU(2) symmetry.\cite{com3} 
 Now, the second equality in eq.~(\ref{eq:F_ij^a}) tells us $F_{ij}^\alpha =0$, 
meaning that the $M$-linear term in eq.~(\ref{eq:s_ij=F_ij}) vanishes. 
 Therefore, the $M$-linear dependence of THC in region 1 is purely a nonperturbative effect. 
 It is originally proportional to $M^3$ but multiplied by the spin diffusion propagator that has $M^2$ 
in the denominator. 
 (Recall that $\tilde {\bm d}_i = (2iM\tau )^{-1} \partial_i {\bm n}$.) 
 This is because the integral in the nonlocal relation (\ref{eq:d}) extends over the scale of 
the spin precession length $\lambda$ (which is equal to the spin decay length in region 1), 
and grows as $M$ is reduced. 
 Note that, because of the condition $q\lambda \ll 1$ ($M \gg Dq^2$), 
the dynamics of the transverse spin density is dominated by precession rather than diffusion.

 Quite recently, a very large THE was found by Vistoli {\it et al.} 
in Ce doped $\rm CaMnO_{3}$ (CCMO) thin films\cite{Bibes}. 
 This material is a weak ferromagnet at low doping (below $\sim 5 \%$),  
and the ferromagnetic moment forms skyrmion bubbles in magnetic field. 
 As the doping rate is reduced towards the Mott transition point, 
the Hall resistivity $\rho_{\rm H} = \sigma_{xy}/\sigma_{xx}^2$ was found to be strongly enhanced; 
more strongly than expected from the strong-coupling (Berry phase) formula, 
$\rho_{\rm H} \propto B_{{\rm s}, z}/n$, which is simply proportional to the inverse carrier density $n$. 
 The authors of Ref.~\cite{Bibes} instead considered that this system is in a  
\lq\lq weaker-coupling regime'' 
and adopt the weak-coupling formula, eq.~(\ref{eq:1}). 
 The result, $\rho_{\rm H} \propto (B_{{\rm s}, z}/n) (M/\epsilon_{\rm F}) 
 \propto B_{{\rm s}, z} mM/(nk_{\rm F}^2) $,  
contains, in particular, the electron mass $m$. 
 By interpreting $m$ to be an effective mass $m^*$ that is enhanced 
as the Mott transition is approached, they reached a good understanding of their experimental results. 
 In this system, the weak-coupling condition ($M\tau < 1$) may be supported by  
the small ferromagnetic moment 
and/or the short $\tau$ due to the proximity to Mott transition, 
but detailed analysis is left to the future. 
 Also, CCMO is an antiferromagnet (with canting) and may require more appropriate analysis. 
 However, it would be reasonable to consider that the THE is governed by the ferromagnetic moment,  
and the above scenario seems to capture the essence of the phenomenon.

 In this letter, we presented a unified description of THE that covers the whole diffusive regime 
from strong to weak coupling. 
 Using the spin gauge field, we considered not only the adiabatic (Berry phase) component 
but also the nonadiabatic component. 
 While the adiabatic Berry phase gives a full account of THE in region 0 (strong-coupling regime), 
it is completely canceled by the nonadiabatic contribution in the weak-coupling regime (regions 1 and 2) 
because of the underlying SU(2) symmetry. 
 Thus the THE in the weak-coupling regime is governed by the nonadiabatic processes 
mediated by the precession (region 1) or precession and diffusion (region 2) of the transverse spin density. 
 The former region may have relevance to the recent experiment on manganite thin films. 
 We found the adiabaticity condition to be $M \tau >1$, in agreement with the literature,\cite{Metalidis}
and the locality condition to be $(q\ell)^2 < M\tau $ or $q\ell_{\rm s} < 1$.

\acknowledgment 

 This work is supported by JSPS KAKENHI Grant Numbers 25400339, 15H05702 and 17H02929. 
This work received support also from the ERC Consolidator Grant \#615759 \lq\lq MINT''.  
KN is supported by Grant-in-Aid for JSPS Research Fellow Grant number 16J05516, and by a Program for Leading Graduate Schools ``Integrative Graduate Education and Research in Green Natural Sciences''.


\begin{thebibliography}{9}
\bibitem{Berry} M. V. Berry, Proc. R. Soc. London A, \textbf{392}, 45 (1984). 
\bibitem{Niu} D. Xiao, M.-C. Chang, and Q. Niu, Rev.~Mod.~Phys.  {\bf 82}, 1959 (2010). 
\bibitem{Ye} J. Ye, Y. B. Kim, A. J. Millis, B. I. Shraiman, P. Majumdar, and Z. Te\v{s}anovi\`c, Phys. Rev. Lett. \textbf{83}, 3737 (1999).
\bibitem{Bruno} P. Bruno, V. K. Dugaev, and M. Taillfumier, Phys. Rev. Lett. \textbf{93}, 096806 (2004).
\bibitem{Tatara} G. Tatara and H. Kawamura, J. Phys. Soc. Jpn. \textbf{71} 2613 (2002).
\bibitem{Nakazawa1} K. Nakazawa and H. Kohno, J. Phys. Soc. Jpn. \textbf{83}, 073707 (2014).
\bibitem{Stern} A. Stern, Phys. Rev. Lett. \textbf{68} 1022 (1992).
\bibitem{LSG1} D. Loss, H. Schoeller, and P. M. Goldbart, Phys. Rev. B \textbf{48}, 15218 (1993).
\bibitem{LKPB} S. A. van Langen, H. P. A. Knops, J. C. J. Paasschens, and C. W. J. Beenakker, Phys. Rev. B \textbf{59}, 2102 (1999).
\bibitem{LSG2} D. Loss, H. Schoeller, and P. M. Goldbart, Phys. Rev. B \textbf{59}, 13328 (1999).
\bibitem{Metalidis} G. Metalidis and P. Bruno, Phys. Rev. B \textbf{74}, 045327 (2006). 
\bibitem{Korenman} V. Korenman, J. L. Murray, and R. E. Prange, Phys. Rev. B \textbf{16}, 4032 (1977).
\bibitem{Volovik} G. E. Volovik, J. Phys. C \textbf{20}, (1987).
\bibitem{Bazaliy} Ya. B. Bazaliy, B. A. Jones, and S.-C. Zhang, Phys. Rev. B \textbf{57}, R3213 (1998).
\bibitem{Kohno} H. Kohno and J. Shibata, J. Phys. Soc. Jpn. \textbf{76}, 063710 (2007).
\bibitem{Nakazawa2} K. Nakazawa and H. Kohno, in preparation.
\bibitem{com_A} $A_\mu^\alpha$ in the present letter corresponds to $2A_\mu^\alpha$ in Refs.~\cite{Kohno,Shibata} 
\bibitem{Shibata} J. Shibata and H. Kohno, Phys. Rev. B \textbf{84}, 184408 (2011). 
\bibitem{SM} (Supplemental material) Calculations of 
eqs.~(\ref{eq:s_xy^z0_R0}), (\ref{eq:s_xy^p_R0}), (\ref{eq:s_xy^z0}), (\ref{eq:s_xy^p0}), (\ref{eq:s_xy^pM}) 
and some others, together with all diagrams, are provided online. 
\bibitem{Fukuyama} Calculation of the ${\bm A}^z$-term is the same as the ordinary Hall effect, 
see H. Fukuyama, H. Ebisawa, and Y. Wada, Prog. Theor. Phys. {\bf 42}, 494 (1961). 
\bibitem{com1} This procedure was used in Ref.~\cite{Nakazawa2} to discuss the relation 
of the results obtained by different methods. 
\bibitem{com2} 
  Note that both $\sigma_{ij}^{\rm H}$ and $F_{ij}^\alpha$ are antisymmetric in the real-space indices, 
$i$ and $j$, and that $\alpha$ in $F_{ij}^\alpha$ is a spin index. 
\bibitem{com3} 
 In the presence of $M$, which breaks the rotation symmetry in spin space, 
$A_i^z$ and ${\bm A}_i^\perp$ are not equivalent and the gauge symmetry 
is reduced from SU(2) to U(1). 
 Therefore, instead of eq.~(\ref{eq:s_ij=F_ij}), one would expect generally 
\begin{align}
 \sigma_{ij}^{\rm H} = c_1(M)  (\partial_i  A_j^\alpha - \partial_j A_i^\alpha ) M^\alpha 
               - c_2(M) ({\bm A}_i \times {\bm A}_j)^\alpha M^\alpha + \cdots , 
\nonumber 
\label{eq:s_ij=F_ij_1}
\end{align}
with two coefficients, $c_1(M)$ and $c_2(M)$. 
 However, the difference $c_1(M) - c_2(M)$ is higher order in $M$ (it starts from $M^2$), 
and as far as the $M$-linear terms are concerned, one has $c_1(0) = c_2(0) \equiv c$, 
leading to eq.~(\ref{eq:s_ij=F_ij}). 
\bibitem{Bibes} L. Vistoli, W. Wang, A. Sander, Q. Zhu, B. Casals, R. Cichelero, A. Barth\'el\'emy, S. Fusil, G. Herranz, K. Nakazawa, H. Kohno, J. Santamaria, W. Wu, V. Garcia and M. Bibes, 
submitted. 


\end{thebibliography}
\end{document}